\documentclass{amsart}
\usepackage{amsfonts}
\usepackage{amsmath}
\usepackage{amssymb}
\usepackage{graphicx}
\usepackage{amscd}

\newtheorem{theorem}{Theorem}
\theoremstyle{plain}

\newtheorem{corollary}{Corollary}

\numberwithin{equation}{section}

\begin{document}
\title[High moments Jarque-Bera tests]{High moments Jarque-Bera tests for
arbitrary distribution functions}

\author{Gane Samb LO$^{*}$}
\author{Oumar THIAM$^{**}$}
\author{Mohamed Cheikh HAIDARA$^{**}$}
\address{$^{*}$ LSTA, Universit\'e Pierre et Marie Curie, France and LERSTAD, Universit\'e Gaston Berger
de Saint-Louis, SENEGAL\\
gane-samb.lo@ugb.edu.sn, ganesamblo@ganesamblo.net}
\address{$^{**}$ LERSTAD, Universit\'e Gaston Berger de Saint-Louis, SENEGAL%
\\
ehdeme@ufrsat.org}

\begin{abstract}
The Jarque-Bera's fitting test for normality is a celebrated and powerful one. In this paper, we consider general Jarque-Bera tests for
any distribution function (\textit{df}) having at least \textit{4k} finite
moments for $k \geq 2$. The tests use as many moments as possible whereas
the JB classical test is supposed to test only skewness and kurtosis for
normal variates. But our results unveil the relations between the
coeffients in the JB classical test and the moments, showing that it really
depends on the first eight moments. This is a new explanation for the
powerfulness of such tests. General Chi-square tests for an arbitraty
model, not only normal, are also derived. We make use of  the modern functional empirical processes
approach that makes it easier to handle statistics based on the high moments and
allows the generalization of the JB test both in the number of involved moments and in the underlying distribution. Simulation studies are provided and comparison cases with the
Kolmogorov-Smirnov's tests and the classical JB test are given.
\end{abstract}

\keywords{Asymptotic distribution, Asymptotic statistical tests, normality tests, functional empirical processes}

\subjclass[2000]{Primary 62E20, 62F05. Secondary 62F121, 62GF17}
\maketitle

\Large

\section{Introduction}

\noindent In this paper, we are concerned with generalizations of Jarque-Bera's (JB) 
\cite{jarque} tests based on arbitrary first $(4k)$ moments, $k\geq 2$,
rather than on the first eight ones as usual. (See \cite{neil} for a
reminder of JB tests, page 69). We obtain general statistics that allow
 statistical tests for any distribution function $G$\ provided it has enough
moments. For a reminder, the classical \textit{JB} test belongs to the class
of omnibus moment tests, i.e. those which assess simultaneously whether the
skewness and kurtosis of the data are consistent with a Gaussian model. This
test proved \textit{optimum asymptotic power and good finite sample
properties} (see \cite{jarque}). A detailed description of that test and related indepth analyses can be found in Bowman and Shenton, D'Agosto, D'Agostino et \textit{al.}, etc. (See \cite%
{bowman}, \cite{dagostino}, \cite{dagostinopers} and \cite{dagostinotiet}).%
\newline

\noindent Let $X$, $X_{1}$, $X_{2}$, ... be a sequence of independent and identically distributed random variables (\textit{r.v.}'s) defined on the same probability space 
$(\Omega, \mathcal{A}, \mathbb{P})$. For each $n \geq 1$, the skewness and kurtosis coefficients related to the sample $X, ..., X_{n}$ are defined by

\begin{equation}
b_{n,2}=\frac{\left( 1/n\right) \sum\limits_{i=1}^{n}\left( X_{1}-\overline{X%
}\right) ^{3}}{\left[ \left( 1/n\right) \sum\limits_{i=1}^{n}\left( X_{1}-%
\overline{X}\right) ^{2}\right] ^{3/2}};a_{n,2}=\frac{\left( 1/n\right)
\sum\limits_{i=1}^{n}\left( X_{1}-\overline{X}\right) ^{4}}{\left[ \left(
1/n\right) \sum\limits_{i=1}^{n}\left( X_{1}-\overline{X}\right) ^{2}\right]
^{2}}.  \label{01}
\end{equation}%
\bigskip

\noindent These statistics are designed to estimate the theoretical skewness
and kurtosis given by $b_{2}=\mathbb{E}\left( X-m\right) ^{3}/\sigma ^{3}$
and $a_{2}=\mathbb{E}\left( X-m \right) ^{4}/\sigma ^{4}$ where $m =\mathbb{E%
}(X)$ and $\sigma ^{2}=var(X)$ respectively denote the mean and the variance
of $X$ that is supposed to be nondegenerated. Here and in all the sequel, $\mathbb{E}$ stands for the mathematical expectation with respect to the probability $\mathbb{P}$. Now, under the hypothesis :\newline

\noindent \textit{H0} : $X$ follows a Gaussian normal law,\newline

\noindent we have $b^{2}=0$ and $a=3$ and the \textit{JB} statistic 
\begin{equation}
T_{n}=\frac{n}{6}\left( b_{n,2}^{2}+\frac{1}{4}\left( a_{n,2}-3\right)
^{2}\right)  \label{jb}
\end{equation}

\noindent has an asymptotic chi-square distribution with two degrees of
freedom under the null hypothesis of normality. Jarque-Bera's test consists
in rejecting \textit{H0} when $T_{n}$ is far from zero. We will find below
that the constants $6$ and $24$ used in (\ref{jb}), actually, are closely
related to the first four even moments of a $\mathcal{N}(0,1)$ random
variable which are $1,3,15$ $and$ $105$ and a more convenient form of (\ref%
{jb}) is 
\begin{equation*}
T_{n}=n\left( b_{n,2}^{2}/6+\left( a_{n,2}-3\right)
^{2}/24\right) .
\end{equation*}

\bigskip

\noindent Our objective here is to generalize \textit{JB}'s test to a
general \textit{df}\ $G$ by considering high \ moments $m_{\ell }=\mathbb{E}%
(X^{\ell }),$ $\ell \geq 1,$ with $m_{1} \equiv m$,instead of the first eight
moments only. We base our methods on the remark that for a random variable $%
X\sim \mathcal{N}(m,\sigma ^{2}),$ one has

\begin{equation}
\forall k\geq 0, \mathbb{E}\left( \left( X-m \right) ^{2k+1}\right) =0,\text{
}\mathbb{E}\left( \left( X-m \right) ^{2k}\right) =\frac{\left( 2k\right) !}{%
2^{k}k!}\sigma ^{2k}.  \tag{H1}
\end{equation}

\noindent Actually \textit{JB}'s test only checks the third and fourth
moments of $X$ while the coefficients of the \textit{JB} statistic (\ref{jb}) uses the first eight moments of $X$. Our guess is that we would
have better tests if we were able to simultaneously check all the first $(2k)
$ moments for some $k\geq 2$. To this purpose, we consider the following
statistics, that is the normalized centered empirical moments (NCEM), 
\begin{equation}
b_{n,p}=\frac{\mu _{n,2p-1}}{\mu _{n,2}^{\left( 2p-1\right) /2}}\text{ and }%
a_{n,p}=\frac{\mu _{n,2p}}{\mu _{n,2}^{p}},\text{ }p\geq 2,  \label{momemp}
\end{equation}%
where%
\begin{equation*}
m_{n,\ell }=\sum_{i=1}^{n}X_{i}^{\ell }\text{ and }\mu _{n,\ell }=\frac{1}{n}%
\sum_{i=1}^{n}\left( X_{i}-\overline{X}\right) ^{\ell },\text{ }\ell \geq 1
\end{equation*}

\noindent are the $\ell ^{th}$ non-centered and the centered empirical moments.
By the classical law of large numbers, the statistics in (\ref{momemp}) are,
for each fixed $p$, asymptotic estimators of

\begin{equation}
b_{p}=\frac{\mathbb{E}\left( (X-m)^{2p-1}\right) }{\sigma ^{(2p-1)}}\text{ and }a_{p}=%
\frac{E\left( (X-m)^{2p}\right) }{\sigma ^{2p}},\text{ }p\geq 2,\text{ }
\label{momexa}
\end{equation}%
whenever the $(4p)^{th}$ moment exists. Finally we consider $C^{1}$-class
functions $\left( f_{p}\right) _{p\leq i\leq k}$ et $\left( g_{p}\right)
_{1\leq p\leq k}$ and denote $f=\left( f_{1},...,f_{k}\right) $ and $%
g=\left( g_{1},...,g_{k}\right)$.\\

\noindent  Our general test is based on the
following statistics, for $k\geq 2,$%
\begin{equation}
T_{n}(f,g,k)=\sum_{p=2}^{k}(f_{p}(b_{n,p})+g_{p}(a_{n,p})),  \label{02}
\end{equation}

\noindent which almost-surely $(a.s)$ tends to  
\begin{equation}
T(f,g,k)=\sum_{p=2}^{k}(f_{p}(b_{p})+g_{p}(a_{p})),  \label{02b}
\end{equation}

\noindent as $n\rightarrow +\infty$. For an independent and identically distributed sequence $X_{1}$, $%
X_{2}$, ... of \textit{r.v.}'s associated with a distribution function $G$
having a finite $2k$-moment, we will have by Theorem \ref{theo1} below that 
\begin{equation*}
T_{n}(f,g,k)-T(f,g,k)\overset{\mathbb{P}}{\rightarrow }0\text{ as }%
n\rightarrow +\infty .
\end{equation*}

\noindent From such a general result, we are able to derive a normality test
by using it with $b_{p}=0,$ $a_{p}=((2p)/(2^{p}p!)$ for $2\leq p\leq k$, and
rejects normality for a large value of $T_{n}(f,q,k)$.\newline
\qquad

\noindent We are going to establish a general asymptotic normality
of $T_{n}(f,g,k)$ for any $df$'s $G$ with $4k$ finite moments. These results
provide themselves efficient tests for an arbitrary $d.f$. Next, we will derive
chi-square tests that generalize JB's test for higher moments and for
arbitrary \textit{df}'s too.\\

\noindent Our results will show that these tests based on the $%
2k$ moments, need, in fact, the eight $4k$ moments for computing the
variance. This unveils that the classical JB's test is not based only on the
kurtosis and the skewness but also on the sixth and the eighth moments. To
describe the complete form of the Jarque-Bera method, put \newline

\begin{equation*}
\text{ }aj(p)=\sigma ^{-(4p)}E(X^{2p}-pE(X^{2p})X^{2})^{2}\text{ and }%
bj(p)=\sigma ^{-(4p-2)}E(X^{4p-2}).
\end{equation*}

\bigskip \noindent The JB's test for a $\mathcal{N}(m,\sigma ^{2})$ \textit{%
r.v.} will be showed to derive from the following general law 
\begin{equation}
n\left( b_{n,2}-b_{p})^{2}/bj(p)+\left( a_{n,2}-a_{p}\right)
^{2}/aj(p)\right) \sim \chi _{2}^{2}.  \label{jbs}
\end{equation}

\noindent with the particular coefficients $p=2$, $b_{p}=0$ and $a_{p}=3$. This may be a new explanation of the powerfulness of the JB classical tests
since a successful test of normality means that the sample is from a \textit{%
df} having same first eight moments as the $\mathcal{N}(0,1)$ $r.v.$, and this is
very highly improbable for a non normal \textit{r.v.}.\\

\noindent As an illustration of what preceeds, consider a distribution following a double-gamma distribution $\gamma_{d}((1+\sqrt{13})/2,1)$ of density probability $f(x)=b^{a}/(2\Gamma
(a))\left\vert x\right\vert ^{a-1}\exp (-b\left\vert x\right\vert )$ with $a=1+\sqrt(13))/2$. This $rv$ is centered and has a kurtosis coefficient equal to 3. It is rejected from normality by the JB test. If only the skewned and kurtosis do matter, it would not be the case. Actually, the rejection comes from the parameters $aj(2)$ and $bj(2)$ that are very different from a standard normal distribution to this specific distribution.\\

\noindent The rest of the paper is organized as follows. In Subsection \ref{subsec21}
of Section \ref{sec2} we begin to give a concise of reminder the modern
theory of functional empirical processes that is the main theoretical tool
we use for finding the asymptotic law of (\ref{02}). Next in Subsection \ref{subsec22} we establish general results of the consistency of (\ref{02}) and
its asymptotic law, consider particular cases in Subsection \ref{subsec23}%
, propose chi-square universal tests in Subsection \ref{subsec24} and finally state
the proofs in Subsection \ref{subsec25}. We end the paper by Section \ref{sec3}
where simulation results concerning the normal and double-exponential models
are given.\\

\noindent  We here express that in all the sequel, the limits are meant as $%
n\rightarrow +\infty $ and this will not be precised again unless it is
necessary.

\section{RESULTS AND PROOFS}

\label{sec2}

\subsection{A reminder of Functional empirical process}

\label{subsec21}

\noindent Since the empirical functional process is our key tool here, we
are going to make a brief reminder on this process associated with $X_{1}$, $%
X_{2}$, ..., and defined for each $n\geq 1$ by 
\begin{equation*}
\mathbb{G}_{n}(f)=\frac{1}{\sqrt{n}} \sum_{i=1}^{n}(f(X_{i})-\mathbb{E}%
f(X_{i})),
\end{equation*}

\noindent where $f$ is a real measurable function defined on $\mathbb{R}$
such that

\begin{equation}
\mathbb{P}_{G}(\left\vert f\right\vert )=\int \left\vert f(x)\right\vert
dG(x)<\infty \text{ },  \label{esp}
\end{equation}

\bigskip \noindent and

\begin{equation}
\mathbb{V}_{G}(f)=\int \left( f(x)-\mathbb{P}_{G}(f)\right) ^{2}dG(x)<\infty.
\label{var}
\end{equation}

\noindent It is known (see van der Vaart \cite{vaart}, pages 81-93) that $%
\mathbb{G}_{n}$ converges to a functional Gaussian process $\mathbb{G}$ with
covariance function%
\begin{equation}
\Gamma (\overline{f},\overline{\overline{f}})=\int \left( \overline{f}-%
\mathbb{P}_{G}(\overline{f})\right) \left( \overline{\overline{f}}-\mathbb{P}%
_{G}(\overline{\overline{f}})\right) dG(x),  \label{cov}
\end{equation}

\noindent at least in finite distributions. $\mathbb{G}_{n}$ is linear, that is, for $f$ and $g$ satisfying (\ref{var}) and for $(a,b)\in \mathbb{R}
{^{2}}$, we have

\begin{equation*}
a\mathbb{G}_{n}(f)+b\mathbb{G}_{n}(g)=\mathbb{G}_{n}(af+bg).
\end{equation*}

\noindent This linearity will be useful for our proofs. We are now in position to state our
main results.

\subsection{Statements of results}

\label{subsec22}

\noindent First introduce this notation for $\ell \geq 0$, $k\geq 2$, and $%
2\leq p \leq k$. Let $f_{i}$ and $g_{i}$, $i=1,...,k$ be $C^{1}$-functions
with values in $\mathbb{R}$. Put $\mu _{2}=\sigma^{2}$ and $m_{1}=m$ and $%
h_{\ell }(x)=x^{\ell },x\in \mathbb{R}.$

\begin{equation}
A(\ell)=h_{\ell}+ \sum_{p=0}^{\ell-1}C_{\ell}^{p} (-1) ^{\ell-p} \left(
m_{1}^{\ell-p} h_{p} + (\ell-p) m_{1}^{\ell-p-1} m_{p} h_{1} \right)
\label{no1}
\end{equation}

\begin{equation}
B(p)=\sigma^{-(2p-1)}\left( A(2p-1)-\frac{1}{2}(2p-1)\sigma
^{-2}\mu_{2p-1}A(2) \right)  \label{no2}
\end{equation}

\begin{equation}
C(p)=\sigma ^{-2p}\left( A(2p)-p\sigma ^{-2}\mu_{2p}A(2)\right)  \label{no3}
\end{equation}

\bigskip

\noindent and

\begin{equation}
D_{k}=\sum_{p=2}^{k}\left( f_{p}^{\prime }(b_{p})B(p)+g_{p}^{\prime
}(a_{p})C(p)\right) .  \label{no4}
\end{equation}

\noindent Here are our main results.

\begin{theorem}
\label{theo1} Let $\mathbb{E}\left\vert X\right\vert ^{4k}<\infty ,$ for $%
k\geq 2.$ Then%
\begin{equation*}
T_{n}^{\ast }(f,g,k)=\sqrt{n}\left( T_{n}(f,g,k)-T(f,g,k)\right) \rightarrow 
\mathcal{N}\left( 0,\sigma _{k}^{2}\right) ,
\end{equation*}%
\noindent where%
\begin{equation*}
\sigma _{k}^{2}=\left( \int D_{k}^{2}(x)dG(x)\right) -\left( \int
D_{k}(x)dG(x)\right) ^{2}.
\end{equation*}
\end{theorem}

\begin{corollary}
\label{cor1} (Normality test). Let $X$ be a $\mathcal{N}\left(
m,\sigma ^{2}\right) $ \textit{r.v.} and let, for all $k\geq 2$%
\begin{equation*}
T_{k}=\sum_{p=2}^{k}\left( f_{p}(0)+g_{p}\left( \frac{(2p)!}{2^{p}p!}\right)
\right) .
\end{equation*}%
\noindent Then%
\begin{equation*}
\sqrt{n}\left( T_{n}(f,g,k)-T_{k}\right) \rightarrow \mathcal{N}\left(
0,\sigma _{k,0}^{2}\right) ,
\end{equation*}%
\noindent where%
\begin{equation*}
\sigma _{k,0}^{2}=\left( \int D_{k,0}^{2}(x)dG(x)\right) -\left( \int
D_{k,0}(x)dG(x)\right) ^{2},
\end{equation*}%
\noindent and%
\begin{equation*}
D_{k,0}=\sum_{p=2}^{k}\left( f_{p}^{\prime }(0)B(p)+g_{p}^{\prime }\left(
(2p)!/2^{p}p!\right) C(p)\right) .
\end{equation*}
\end{corollary}

\subsection{Particular cases and consequences}

\label{subsec23}

\subsubsection{A general test}

\label{subsec231}

Let $G$ be an arbitrary $df$ with a $4k^{th}$ finite moment for $k\geq 2,$
this is $\int x^{4k}dG(x)<+\infty $. We want to check whether a sample $%
X_{1},..,X_{n}$ is from $G.$ We then select $C^{1}-functions$ $f_{i}$ and $%
g_{i},$ $i=1,...,k$ and compute the observed value $t_{n}^{\ast }(f,g,k)$ of 
$\sqrt(n)(T_{n}^{\ast }(f,g,k)-T^{\ast }(f,g,k))$ and report the $p$-value
of the test, that is $p=\mathbb{P}(\left\vert \mathcal{N}(0,1)\right\vert
\geq \left\vert t_{n}^{\ast }(f,g,k)\right\vert s)$ where $s^2$ is either
the exact variance $\sigma^{2}_{k}$ or its plug-in estimator 
\begin{equation*}
\widehat{\sigma }_{k,n}^{2}=\left( \frac{1}{n}\sum%
\limits_{i=1}^{n}D_{k}^{2}(X_{j,n})\right) -\left( \frac{1}{n}%
\sum\limits_{i=1}^{n}D_{k}(X_{j,n})\right) ^{2}.
\end{equation*}

\noindent Our guess is that using a greater value of $k$ makes the test more
powerful since the equality in distribution of univariate \textit{r.v.}'s
means equality of all moments when they exist (see page 213 in \cite{loeve}%
). For $k=2,$ this result depends on the first eight moments. Then to find
another $df$ $G_{1}$ for which the p-value exceeds $5\%$ would suggest it
has the same eight moments as $G$, which is highly improbable. Simulation
studies in Section \ref{sec3} support our findings. Remark that we have as
many choices as possible for the functions the $f_{i}^{\prime }s$\ and $%
g_{i}^{\prime }s$.\newline

Unfortunately, in the simulation studies reported below, we noticed that the
plug-in estimator $\widehat{\sigma }_{k,n}^{2}$ may hugely over estimate the
exact variance and leads to accepting any data to follow that model, or
significantly underestimate it and leads to reject data form the model
itself. This is why we only use the exact variance here.\newline

\noindent Now let us show how to derive chi-square tests from Theorem \ref%
{theo1}.

\subsubsection{Generalized JB test and tests for symmetrical df's}

\label{subsec232}

Suppose that $X$ is a symmetrical distribution. We have from Theorem \ref%
{theo1} that 
\begin{equation}
\sqrt{n}((b_{n,p}-b_{p}),(a_{n,2}-a_{p}))=(\mathbb{G}_{n}(B(p)),\mathbb{G}%
_{n}(C(p)))+o_{\mathbb{P}}(1).  \label{no5}
\end{equation}%
Since $X$ is symmetrical, that is $\mu _{2\ell -1}=0$ for $\ell \geq 1$, we
may without loss of generality suppose that $m_{1}=0$ since replacing $X$ by 
$X-m_{1}$ does affect neither the $(b_{n,p},a_{n,p})^{\prime }s$ nor the 
$(b_{p},a_{p})^{\prime }s.$ Then we have from (\ref{no1}) and (\ref{no2})
that%
\begin{equation*}
C(p)=\sigma ^{-(2p-1)}A(2p-1)=\sigma ^{-(2p)}(h_{2p}-p\sigma ^{-2}\mu
_{2p}h_{2})
\end{equation*}%
and%
\begin{equation*}
B(p)=\sigma ^{-(2p-1)}(h_{2p-1}-(2p-1)m_{2(p-1)}h_{1}).
\end{equation*}

\noindent By reminding that $h_{p}h_{q}=h_{p+q}$ for $p\geq 0$ and $q \geq 0$, we observe that the product $B(p)\times C(p)$ only includes
functions $h_{j}$ with odd $j^{\prime }s$ and then $\mathbb{E}\mathbb{G}%
_{n}(B(p)\ast C(p))=0$. Thus

\begin{equation*}
\sqrt{n}((b_{n,p}-b_{p}),(a_{n,p}-a_{p}))\rightarrow _{d}\mathbb{N}%
_{2}(0,\Sigma _{p}),
\end{equation*}

\noindent where $\left( \Sigma _{p}\right) _{11}=\mathbb{V}%
ar(B(p))=bj(p),\left( \Sigma _{p}\right) _{22}=\mathbb{V}ar(C(p))=aj(p)$ and 
$\left( \Sigma _{p}\right) _{12}=0$. We get

\begin{corollary}
Let $\int x^{4p}dG(x)<\infty $ for $p\geq 2$ and $G$ be a symmetrical $df$. We have%
\begin{equation}
n(b_{n,p}^{2}{}/bj(p)+(a_{n,p}-a_{p})^{2}/aj(b))\rightarrow \chi _{2}^{2}.
\label{jbG}
\end{equation}
\end{corollary}

\noindent For a standard normal random variable, we get $bj(2)=6$ and $%
aj(2)=24$ and the normality JB's test becomes a particular case of (\ref{jbG}),
which is a general chi-square test for an arbritrary $df$ with $2p$%
-finite moments.

\begin{corollary}
Let $G$ be a Gaussian $df.$ Then%
\begin{equation*}
\frac{n}{6}(b_{n,2}^{2}{}+(a_{n,2}-3)^{2}/4)\rightarrow \chi _{2}^{2}.
\end{equation*}
\end{corollary}

\bigskip

We see that we obtain an infinite number of tests for the normality. For
example, for p=3, we have, $\frac{n}{360}%
(b_{n,3}^{2}{}/2+(a_{n,3}-15)^{2}/17)\rightarrow \chi _{2}^{2},$ etc.

\subsection{A general chi-square test} \label{subsec24}

Consider (\ref{no5}) and put $abj(p)=cov(C(p),B(p))$ and suppose that $%
\Delta (p)=aj(p)\times bj(p)-abj(p)^{2}\neq 0$. We have

\begin{corollary}
Let $\int x^{4k}dG(x)<\infty $ and $\Delta(p) \neq 0$ for $2\leq p \leq k$.
Then%
\begin{equation*}
\frac{n}{\Delta(p)}\left( aj(p)(b_{n,p}-b_{p})^{2})+bj(p)(a_{n,p}-a_{p})^{2}
-2*abj(p)(b_{n,p}-b_{p})(a_{n,p}-a_{p}) \right)
\end{equation*}
\noindent converges in law to a $\chi_{2}^{2}$ $r.v.$.
\end{corollary}

\bigskip

\noindent It is now time to prove Theorem \ref{theo1} before considering the
simulation studies.

\subsection{Proofs}

\label{subsec25}

\noindent Since $G$ has at least first $4k$ moments finite, we are entitled
to use the finite-distribution convergence of the empirical function process 
$\mathbb{G}_{n}$ as below. Let us begin to give the asymptotic law of $%
\mu_{n,\ell}$. By denoting $h_{\ell}(x)=x^{\ell},$ we have

\begin{equation*}
\mu_{n,\ell}=\sum_{p=0}^{\ell}C_{\ell}^{p}\left( -\overline{X}\right)
^{\ell-p}\left( \frac{1}{n}\sum_{i=1}^{n}X_{i}^{p}\right)
\end{equation*}

\begin{equation*}
=\sum_{p=0}^{\ell} C_{\ell}^{p}\left( -1\right) ^{\ell-p}\left(m_{1}+\frac{%
\mathbb{G}_{n}(h_{1})}{\sqrt{n}}\right) ^{\ell-p}\left( m_{p}+\frac{\mathbb{G%
}_{n}(h_{p})}{\sqrt{n}}\right)
\end{equation*}

\begin{equation*}
=\left( m_{\ell}+\frac{\mathbb{G}_{n}(h_{\ell})}{\sqrt{n}}\right) +
\sum_{p=0}^{\ell-1}C_{\ell}^{p}\left( -1\right)
^{\ell-p}\left(m_{1}^{\ell-p}+(\ell-p) m_{1}^{\ell-p-1}\frac{\mathbb{G}%
_{n}(h_{1})}{\sqrt{n}}+o_{p}(n^{-1/2})\right)
\end{equation*}

\begin{equation*}
\times \left( m_{p}+\frac{\mathbb{G}_{n}(h_{p})}{\sqrt{n}}\right)
\end{equation*}

\begin{equation*}
=m_{\ell}+h_{\ell} + \sum_{p=0}^{\ell-1}C_{\ell}^{p} (-1)^{\ell-p} \left(
m_{1}^{\ell-p} m_{p} + \frac{\mathbb{G}_{n}(A_{\ell})}{\sqrt{n}} \right) +
o_{p}(n^{-1/2})
\end{equation*}

\noindent where $A(\ell)$ is defined in (\ref{no1}) and where we used that
the linearity of the empirical functional process. By observing that $\mu_{\ell}=\sum_{p=0}^{\ell}C_{\ell}^{p}\left(-m_{1}\right)
^{\ell-p}\left(m_{p}\right) $, we finally obtain
\begin{equation}
\sqrt{n}\left(\mu_{n,\ell}-\mu_{\ell}\right) =\mathbb{G}_{n}\left(
A(l)\right) +o_{p}(1).  \label{03}
\end{equation}

\noindent Now the law of $b_{n,p}$ is given by

\begin{equation*}
\sqrt{n}\left( b_{n,p}-b_{p}\right) =\frac{1}{\mu _{n,2}^{\left( 2p-1\right)
/2}}\sqrt{n}\left( \mu_{n,2p-1}-\mu_{2p-1}\right)
\end{equation*}

\begin{equation*}
-\frac{\mu_{2p-1}}{\mu _{n,2}^{\left( 2p-1\right) /2}\mu _{2}^{\left(
2p-1\right) /2}}\sqrt{n}\left( \mu _{n,2}^{\left( 2p-1\right) /2}-\mu
_{2}^{\left( 2p-1\right) /2}\right) .
\end{equation*}

\noindent By the delta-method, we have

\begin{equation*}
\mu _{n,2}^{\left( 2p-1\right) /2} =\left( \mu _{2}+\frac{\mathbb{G}%
_{n}(A(2))}{\sqrt{n}}\right) ^{\frac{2p-1}{2}}+ o_{p}(n^{-1/2}).
\end{equation*}

\begin{equation*}
=\mu _{2}^ {\frac{2p-1}{2}}+\frac{2p-1}{2}\mu_{2}^{\frac{2p-3}{2}}\frac{%
\mathbb{G}_{n}(A(2))}{\sqrt(n)}+o_{p}(n^{-1/2}).
\end{equation*}

\noindent and then

\begin{equation*}
\sqrt{n}\left( \mu_{n,2}^{\left( 2p-1\right) /2}-\mu _{2}^{\left(
2p-1\right) /2}\right) =\left( \frac{2p-1}{2}\right) \mu_{2}^{\frac{2p-3}{2}}%
\mathbb{G}_{n}(A(2))+o_{p}(1),
\end{equation*}

\noindent and next, by noticing from \ref{03} that $\mu
_{n,\ell}\rightarrow \mu_{\ell}$ for all $\ell\leq 2k$,

\begin{equation*}
\sqrt{n}\left( b_{n,p}-b_{p}\right)
\end{equation*}

\begin{equation*}
=\mathbb{G}_{n}\left( \sigma ^{-(2p-1)}A(2p-1)-\frac{1}{2}(2p-1)\sigma
^{-(2p+1)}\mu_{2p-1}A(2)\right) +o_{p}(1).
\end{equation*}

\begin{equation*}
\mathbb{G}_{n}\left( B(p)\right) +o_{p}(1)\rightarrow \mathbb{G}\left(
B(p)\right) ,
\end{equation*}

\noindent where $B(p)$ is given in (\ref{no2}). By the very same methods, we
have

\begin{equation*}
\sqrt{n}\left( a_{n,p}-a_{p}\right) =\mathbb{G}_{n}\left( C(p)\right)
+o_{p}(1),
\end{equation*}

\noindent $C(p)$ is stated in (\ref{no3}). The delta-method also yields

\begin{equation*}
\sqrt{n}\left( T_{n}(f,g,k)-T(f,g,k)\right) =T_{n}^{\ast }(f,g,k)
\end{equation*}%
\begin{equation*}
=\sum_{p=2}^{k}\left( f_{p}\left( b_{n,p}\right) -f\left( b_{p}\right)
\right) +\sum_{p=2}^{k}\left( g_{p}\left( a_{n,p}\right) -g\left(
a_{p}\right) \right)
\end{equation*}

\begin{equation*}
=\sum_{p=2}^{k}f_{p}^{\prime }(b_{p})\mathbb{G}_{n}\left( B(p)\right)
+\sum_{p=2}^{k}g_{p}^{\prime }(a_{p})\mathbb{G}_{n}\left( C(p)\right)
+o_{p}(1)
\end{equation*}

\begin{equation*}
=\mathbb{G}_{n} \biggl( \sum_{p=2}^{k}\left(f_{p}^{\prime }(b_{p}) B(p)
+g_{p}^{\prime }(a_{p}) C(p)\right)\biggr) +o_{p}(1)
\end{equation*}

\begin{equation*}
=\mathbb{G}(D_{k})+o_{p}(1).
\end{equation*}

\noindent This completes the proof of the theorem. The proof of the corollary
is a simple consequence of the theorem.

\section{Simulation and Applications}

\label{sec3}

\subsection{Scope the study} \label{subsec31}

We want to focus on illustratring how performs the general test for usual
laws such as Normal and Double Gamma ones. It is clear that the generality
of our results that are applicable to arbitrary $d.f.$'s with some finite $%
k^{th}$-moment $(k\geq 2)$ deserves extended simulation studies for
different classes of $df$'s. We particularly have to pay attention
to the choice of $k$ and of the functions $f_{i}$ and $g_{i}$, depending on
the specific model we want to test.\newline

\noindent In this paper, we want to set a general and workable method to simulate and
test two symmetrical models. The normal and the double-exponential one with
density $f(x)=(\lambda /2)\exp (-\lambda \left\vert x\right\vert )$. We expect to find a test that accepts normality for normal data and
rejects double-exponental data and to confirm this by the Jarque-Berra test,
and to have an other test that exactely does the contrary.\newline

Once these results are achieved, we would be in position to handle a larger
scale simulation research following the outlined method. Specially, fitting
financial data to the generalized hyberpoblic model is one the most
interesting applications of our results.

\bigskip

\subsection{The frame} \label{subsec32} We first choose all the functions $f_{i}$ equal to $f_{0}$ and all the
functions $g_{i}$ equal to $g_{0}$. We fix $k=3$, that is we work with the
first twelve moments. As a general method, we consider two $df^{\prime }s$ $%
G_{1}$ and $G_{2}.$ We fix one of them say $G_{1}$ and compute $%
T(f,g,k)=T(f,g,k,G_{1})$ and the variance $\sigma _{k}^{2}$ from the exact
distribution function $G_{1}$. We generate samples of size $n$ from one
the $df^{\prime }s$ (either $G_{1}$ or $G_{2})$ and compute $T_{n}(f,g,k).$
We repeat this $B$ times and report the mean value $t^{\ast } $ of the
replicated values of $T_{n}^{\ast }=\sqrt{n}\left(
T_{n}(f,g,k)-T(f,g,k)\right) /\sigma $ and report the p-value $p=\mathbb{P}%
(\left\vert \mathcal{N}(0,1)\right\vert \geq t^{\ast })$. The simulation
outcomes will be considered as conclusive if $p$ is high for samples from $%
G_{1}$ and low for samples from $G_{2}$. The results are compared with those given by the Kolmogorov-Smirnov test (KST) and when the data are Gaussian,
they are compared with the outcomes from JB's classical test.

\subsection{The results}

 We consider the following cases : $G_{1}$ is a Gaussian 
\textit{r.v} $\mathcal{N}(m,\sigma^{2})$; $G_{2}$ is double-exponential law
\ $\mathcal{E}_{d}(\lambda )$\ with density probability $f_{2}(x)=(\lambda
/2)\exp (-\lambda \left\vert x\right\vert )$ and $G_{3}$ is a double-gamma
law $\gamma _{d}(a,b)$ with probability density $f_{3}(x)=b^{a}/(2\Gamma
(a))\left\vert x\right\vert ^{a-1}\exp (-b\left\vert x\right\vert )$.

\subsubsection{Normal Model N(m,$\protect\sigma^{2})$}

The choice $f_{0}(x)=g_{0}(x)=x^{2}$ is natural since the Jarque-Berra test
may be derived for our result for these functions and for $k=2.$ The model
is determined by these following parameters :

\begin{equation*}
\begin{tabular}{|l|l|l|}
\hline
$(b_{p},a_{p}),2\leq p\leq 6$ & $T(f,g,k)$ & $\sigma $ \\ 
\hline
$(0,3),(0,15),(0,105),(0,946),(0,10395)$ & $234$ & $500.2918$\\
\hline
\end{tabular}
%\caption{Table 1 : Parameters of the  model.}
\end{equation*}

\noindent We recall that the variance of our statistic depends on the first $4k$ moments.\newline

\noindent \textbf{Simulation study}.

Testing the model with $\mathcal{N}(0,1)$ data gives the following outcomes
for $n=20$

\begin{equation*}
\begin{array}{|c|c|c|c|c|c|c|c|}
\hline
& T_{n}(f,g,k) & T_{n}^{\ast } & p\% & JB  & pJB \% & KS  & pKS \%\\
\hline
N(0,1) & 232.16 & -0.023 & 49.05 & 1.338 & 51.5 & 0.7709 & 23.35 \\ 
\hline
\end{array}
%\caption{Table 2 : Simulation $mathcal{N}(00,1)$ model for small samples $n=20$}
\end{equation*}

\noindent and for $n=100$,
\begin{equation*}
\begin{array}{|c|c|c|c|c|c|c|c|}
\hline
& T_{n}(f,g,k) & T_{n}^{\ast } & p\% & JB  & pJB \% & KS  & pKS \%\\
\hline
N(0,1) & 249.21 & 0.42 & 33.82 & 1.73 & 42.22 & 0.918 & 15.60 \\ 
\hline
\end{array}
%\caption{Table 2 : Simulation $mathcal{N}(00,1)$ model for small samples $n=20$}
\end{equation*}

\noindent and for $n=1000$,

\begin{equation*}
\begin{array}{|c|c|c|c|c|c|c|c|}
\hline
& T_{n}(f,g,k) & T_{n}^{\ast } & p\% & JB  & pJB \% & KS  & pKS \%\\
\hline
N(0,1) & 243.34 & 0.59 & 27.73 & 2.08 & 35.38 & 0.98 & 12.62 \\ 
\hline
\end{array}
%\caption{Table 2 : Simulation $mathcal{N}(00,1)$ model for small samples $n=20$}
\end{equation*}

\noindent where $JB$ is the classical Jarque-Berra statistic, $pJB$ is
the p-value of the JB test, $KS$ is the Kolmogorov-smirnov statistic and $pKS$ is the related p-value. Our model accepts the normality and this is
confirmed by JB's test and by the Klmogorov-Smirnov test (KST). The simulation results are very stable and
constantly suggest acceptance.\newline

\noindent \textbf{Testing the double-exponential versus the normal model}

Recall that the values $(b_{p},a_{p})$ for $2\leq p\leq 6$ $\ $are $%
(0,3),(0,15),(0,946),(0,10395)$. Comparing these values with those of a normal
model, it is natural to think that the test will fail since only the $%
b_{p}$ coincide and the test is only based on the moments. Indeed, using
data from $\mathcal{E}_{d}(1)$ gives for $n=11$

\begin{equation*}
\begin{array}{|c|c|c|c|c|c|c|c|}
\hline
& T_{n}(f,g,k) & T_{n}^{\ast } & p \% & JB  & pJB\% & KS & pKS\%\\ 
\hline
\mathcal{E}(1) & 411.25 & 1.81 & 3.47 & 1.98 & 37.98 & 0.91 & 15.67\\ 
\hline
\end{array}
%\caption{Table 2 : testing $mathcal{N}(00,1)$ model with double exponential data for small samples $n=20$}
\end{equation*}

\noindent and for $n=22$\newline

\begin{equation*}
\begin{array}{|c|c|c|c|c|c|c|c|}
\hline
& T_{n}(f,g,k) & T_{n}^{\ast } & p\% & JB & pJB \% & KS & pKS\%\\ 
\hline
\mathcal{E}(1) & 1624 & 18.70 & 0 & 6.43 & 4.09 & 0.9 & 15\\ 
\hline 
\end{array}
%\caption{Table 2 : testing $mathcal{N}(00,1)$ model with double exponential data for small samples $n=20$}
\end{equation*}

\noindent Our test rejects the $\mathcal{E}_{d}(1)$ model for $n=11$ and JB's test rejects it only for $n \geq 22$. We see here the advantage brought by the value $k=3$ in our statistic.
The KST has problems in rejecting the false $\mathcal{E}_{d}(1)$ even for $n=1000$
that of Jarque-Berra.\newline

\noindent \textbf{Testing the double-gamma versus the normal model}.\newline

\noindent Let use $\gamma _{d}(a,b)$ data with $a_{0}=(1+\sqrt{13})/2$ and $%
b=1$. We have the outcomes for $n=11$

\begin{equation*}
\begin{array}{|c|c|c|c|c|c|c|c|}
\hline
& T_{n}(f,g,k) & T_{n}^{\ast } & p \% & JB  & pJB\% & KS & pKS\%\\ 
\hline
\mathcal{E}(1) & 527.8 & 3.09 & 0.099 & 4.22 & 12.5 & 0.99 & 12.45\\ 
\hline
\end{array}
%\caption{Table 2 : testing $mathcal{N}(00,1)$ model with double exponential data for small samples $n=20$}
\end{equation*}

\noindent and for $n=22$\newline

\begin{equation*}
\begin{array}{|c|c|c|c|c|c|c|c|}
\hline
& T_{n}(f,g,k) & T_{n}^{\ast } & p\% & JB & pJB \% & KS & pKS\%\\ 
\hline
\mathcal{E}(1) & 1055 & 10.16 & 0 & 6.41 & 4.12 & 0.99 & 11\\ 
\hline 
\end{array}
%\caption{Table 2 : testing $mathcal{N}(00,1)$ model with double exponential data for small samples $n=20$}
\end{equation*}

\noindent We have similar results. Ou test rejects the $\mathcal{E}_{d}(1)$ model for $n=12$ and JB's test rejects it only for $n \geq 18$. We see here the advantage brought by the value $k=3$ in our statistic. Although the first four moments of a $\gamma _{d}(a_{0},1)$ are $%
0,1,0$ and $3$, that is, the same of those of standard normal $rv$, this
model is rejected. We already pointed out that the coefficients $4$ and $6$
are in fact based on the first eight moments and the discrepancy of moments
higher than $4$ results in the rejection.\newline

\noindent Analysing the tables above, we conclude that our test performs better the JB's test against a double-gamma \textit{df} with same skewness and
kurtosis than a normal \textit{df} for small sample sizes around ten and this is real advantage for small data sizes. Even for $k=2$, our test is performant for the small values
$n=11$ and $n=12$.\\

\noindent \textbf{Double-exponential model $\mathcal{E}_{d}$($\lambda )$}.%
\newline

\noindent We point out that the statistic $T_{n}(f,g,k)$ does not depend on the $\lambda$. Then we only consider $\lambda=1$ in the following. 
\noindent We always use $f_{0}(x)=g_{0}(x)=x^{2}.$ The model is determined by the following values.

\begin{equation*}
\begin{tabular}{|l|l|l|}
\hline
$(b_{p},a_{p}),2\leq p\leq 6$ & $T(f,g,k)$ & $\sigma $ \\ 
\hline
$(0,6),(0,90),(0,2520),(0,113400),(0,7484400)$ & $8136$ & $73473$\\
\hline
\end{tabular}%
\end{equation*}

\noindent Here, we do not have the Jarque-Berra test to confirm the results.
\newline

\noindent \textbf{Simulation}. Testing the model with $\mathcal{E}_{d}$($%
\lambda )$ data gives the following outcomes, for $n=800$.

\begin{equation*}
\begin{array}{|c|c|c|c|}
\hline
& T_{n}(f,g,k) & T_{n}^{\ast } & p\% \\
\hline 
\mathcal{E}_{d}(1) & 7858,0174 & -0.41 & 41,370\\
\hline
\end{array}%
\end{equation*}

\noindent The simulation results are very stable and constantly suggest
acceptance.\\

\noindent \textbf{Testing normal data}. Using normal data gives

\begin{equation*}
\begin{array}{|c|c|c|c|}
\hline
& T_{n}(f,g,k) & T_{n}^{\ast } & p\% \\ 
\hline
\mathcal{N}(0,1) & 236.019& -3.044 & 0.11\\
\hline
\end{array}%
\end{equation*}

The $\mathcal{N}(0,1)$ model is rejected. We noticed that the rejection of normal data is automatically obtained for large sizes here, when $n$ is greater than $900$. For $n$
between $500$ and $900$, rejection is frequent but acceptance occurs now and
then. Whe also noticed that the variance of $T_{n}^{\ast }$ are high and do not
allow to reject normal data for small sizes. This leads us to consider other
functions. Now consider the classes of functions
\begin{equation*}
\theta u+(1+u^{p})^{p},p\text{ even.}
\end{equation*}%
We obtain good results for $n=150$ with $\theta =0.1$ and $p=2.$ In this
case, the exact value of the statistic is $11.600$. The double-exponential 
$\mathcal{E}_{d}$($1)$ model is confirmed according to the following table

\begin{equation*}
\begin{array}{|c|c|c|c|}
\hline
& T_{n}(f,g,k) & T_{n}^{\ast } & p\% \\ 
\hline
\mathcal{E}_{d}(1) & 7.968 & -0.7973 & 21.38\\
\hline
\end{array}
\end{equation*}

\noindent while the normal model is rejected as illustrated below :

\begin{equation*}
\begin{array}{|c|c|c|c|}
\hline
& T_{n}(f,g,k) & T_{n}^{\ast } & p\% \\ 
\hline
\mathcal{N}(0,1) & 3.001 & -1.87 & 3.01\\
\hline
\end{array}
\end{equation*}
\newline

\noindent It is important to mention here that the KST is very powerfull is rejecting the normal model with double-exponential and double-gamma data with extremely low p-value's.

\subsection{Conclusion and perspectives}

We proposed a general test for an arbitrary model. The methods are based on
functional empirical processes theory that readily provided asymptotic laws
from which statistical tests are derived. They depend on an integer $k$ such
that the pertaining $df$ has $4k$ first $finite$ moments. We got two kinds of
tests. A general one based on functions $f_{i\text{ }}$and $g_{i}$, $i=1,...,k$,
with an asymptotic normal law. We derived from these results $chi-square$
tests that are valid for general \textit{df}'s and that includes the Jarque-Berra test of normality. 
Both tests used arbitrary moments. We only undergone simulation
studies for the first kind of test. Our simulation studies showed high
performance for normality against other symmetrical laws such as
double-exponential or double-gamma ones. For suitable choices of $f_i$, $g_i$
and $k,$ the test performs well for small samples $(n=20)$ both for
accepting the normal model and rejecting other models. We also showed that
for suitable choice of $f_i$ and $g_i$, the test for the double-exponential model is
also successful, but for sizes greater that $n=150$. In upcoming papers, we
will focus on detailed results on specific models and try to found out, for
each case, suitable value of the parameters of the tests ensuring good
performances for small data. A paper is also to be devoted to simulation
studies for the $khi-square$ tests and their applications to financial
data.

\end{document}